\documentclass[prb,twocolumn]{revtex4}
\usepackage{amsmath,amssymb,graphicx}
\begin{document}
\title{Interfacial Magnetoelectric Coupling in Tri-component Superlattices} 
\author{Jaekwang Lee}
\author{Na Sai}
\email[]{nsai@physics.utexas.edu}
\author{Tianyi Cai}
\author{Qian Niu}
\author{Alexander A. Demkov}
\email[]{demkov@physics.utexas.edu}
\affiliation{Department of Physics, The University of Texas at Austin, Austin, Texas, 78712}
\begin{abstract}
Using first-principles density functional theory, we investigate the interfacial magnetoelectric coupling in a tri-component superlattice composed of a ferromagnetic metal (FM), ferroelectric (FE), and normal metal (NM). Using Fe/FE/Pt  as a model system, we show that a net and cumulative interfacial magnetization is induced in the FM metal near the FM/FE interface. A carefully analysis of the magnetic moments in Fe reveals that the interfacial magnetization is a consequence of a complex interplay of interfacial charge transfer, chemical bonding, and spin dependent electrostatic screening. The last effect is linear in the FE polarization, is switchable upon its reversal, and yields a substantial interfacial magnetoelectric coupling. 
\end{abstract}
\maketitle

Materials with coupled magnetic and electric degrees of freedom are classified as magnetoelectrics. A strong magnetoelectric coupling enables control of magnetism by the electric field or vice versa, and hence has a strong appeal for  emerging device applications.~\cite{erenstein, ramesh, cheong, scott}  A linear coupling between these two degrees of freedom involves breaking both space and time reversal symmetries and is therefore of fundamental interest. Multiferroic oxides in which simultaneous ferroelectric and magnetic orders exist form a class of promising magnetoelectric materials. However, many single phase multiferroics have rather limited application either because of their low Curie temperature or weak coupling.

Heterointerfaces have proven to be ideal for controlling and manipulating electrical charges and spins in solid state devices.  
Recently, efforts have focused on the so called ``interfacial magnetoelectricity'' in which a magnetoelectric coupling  arises at a ferromagnetic metal/dielectric (or ferroelectric) interface.~\cite{zhang, rond, duan06, cai, duan08,Fechner,Niranjan, Niranjan2} A ferroelectric or a dielectric (upon the application of an electric field) can induce free charges near the interface within a ferromagnetic metal. These screening charges are spin dependent in the ferromagnet, and thus in turn yield an additional magnetization that exists only within a nanometer near the interface. 
In contrast to the intrinsic coupling in  multiferroic oxides, this effect can be viewed as an extrinsic magnetoelectric effect and may offer an alternative to a single phase multiferroic in providing the robust room temperature magnetoelectric (ME) effect. 
\begin{figure}
\includegraphics*[width=8cm]{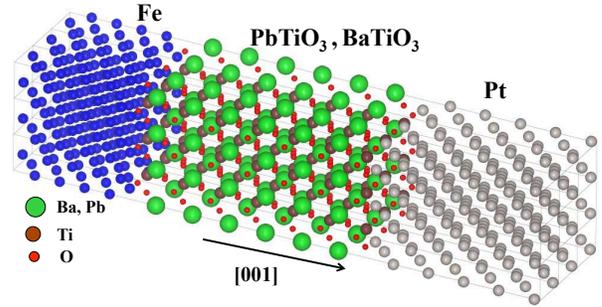}
\caption{Schematics of the ``tri-component'' superlattices which consist of alternatively stacked ferromagnetic metal (Fe), ferroelectric, and normal metal (Pt). }   
\label{structure}
\end{figure}

In this paper we explore ``tri-component'' FM/FE/NM superlattices consisting of alternating layers of a ferromagnetic metal, ferroelectric, and normal metal as depicted in Fig.~\ref{structure}. Unlike the symmetric heterostructure,~\cite{duan06,rond,Fechner,Niranjan, Niranjan2} the broken inversion symmetry in an asymmetric superlattice permits a leading magnetoelectric coupling in the free energy expansion that is linear in the electric polarization of the FE,~\cite{cai}  i.e., $\propto PM^2$, and hence can be electrically controlled.  Moreover, since the induced magnetization only occurs at the FM/FE interface, the overall induced magnetization
does not cancel as in the symmetric structure\cite{rond} and can cumulate in a superlattice to achieve a large macroscopic magnetization.

Using density functional theory (DFT), we demonstrate this effect using Fe/FE/Pt superlattices, where for FE we use BaTiO$_{3}$ (BTO) and PbTiO$_3$ (PTO) as two examples.  PTO and BTO are two prototypical FEs with robust room temperature polarization. Their significant difference in polarization allows us to investigate the polarization dependence of the ME coupling effect.  A rather complex picture emerges as a result of a detailed electronic structure analysis. The interface magnetization is controlled by the interplay of quite different physical effects including electrostatic screening, contact potential difference, formation of the Fe-O chemical bonds, metal induced gap states, and subtle chemical differences resulting from the polarization switching. The electrostatic effect is approximately linear in
the FE polarization and switchable upon the reversal of the polarization, in agreement with the analytical model.~\cite{cai} 
The overall magnetoelectric coupling we find in these systems is remarkably large compared to many other existing composite structures that have been reported so far.\cite{zhe04,zav05} 

We use DFT within the local density approximation (LDA) as implemented in the VASP code and projector augmented-wave pseudopotentials.~\cite{VASP}  We apply a plane wave cut off energy of 600 eV and the 8$\times$8$\times$8 and 8$\times$8$\times$2 k-point meshes for the Brillouin zone integration for the bulk and superlattice, respectively. Calculations of the lattice parameter in the tetragonal phase yield $3.86$\AA, $c/a=1.04$ for PTO and $3.95$~\AA, $c/a=1.01$ for BTO. The lattice constants are about $1.5\%$ less than the experimental values of $3.904$ and $3.994$\AA,~\cite{exp} as is typical for DFT-LDA calculations. For two metal electrodes, {\it bcc} Fe and {\it fcc} Pt, we find lattice parameters of 3.904 and 3.901~\AA, respectively, to be compared to the experimental values of 4.05 and 3.92~\AA.  We terminate the FEs with TiO$_2$ planes and place the Fe and Pt atoms atop oxygen. For each of the superlattices, (Fe)$_{11}$(PTO)$_9$(Pt)$_8$ and (Fe)$_{9}$(BTO)$_{13}$(Pt)$_{10}$ (the subscripts indicate the number of atomic layers), we start with the FE displacements calculated for the bulk phase and fully relax the structure until the maximum force is below 10 meV/$\rm \AA$, keeping the in-plane lattice constants fixed to that of the corresponding bulk tetragonal phase of the ferroelectric. 
In the case of BTO, this constraint leads to in-plane tensile strain of about $1\%$ and 1.2$\%$ in Fe and Pt, respectively, and in the case of PTO compressive strain of about $-1\%$ and $-0.9\%$ in Fe and Pt, respectively.~\cite{note_strain}

\begin{figure}
\includegraphics*[width=8cm]{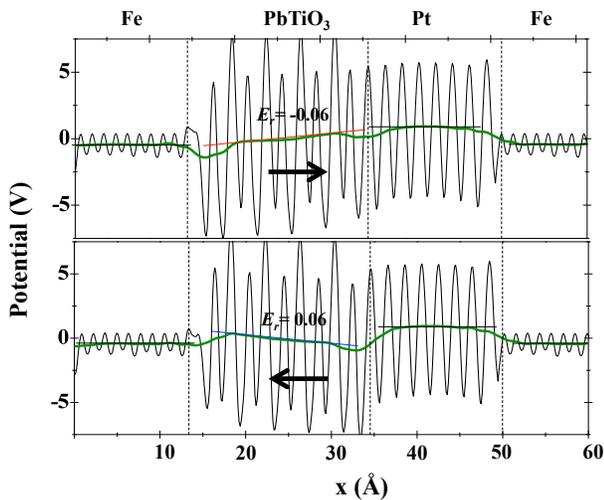}
\caption{The planar (thin gray curve) and macroscopically averaged potential (thick green curve) as a function of distance along the [001] direction for the fully relaxed $\rm(Fe)_{11}(PbTiO_{3})_{9} (Pt)_{8}$ superlattice. The straight line in PTO marks the residual field in the PTO unit cells, where an magnitude of $\pm0.06$ eV/\AA ~ is calculated. The arrows mark the direction of the FE polarization in PbTiO$_3$.}   
\label{pot_pto}
\end{figure}
 We start with discussing the Fe/PbTiO$_3$/Pt superlattice.  The macroscopically averaged electrostatic potentials of the relaxed supercell for two opposite directions of FE polarization are shown in Fig.~\ref{pot_pto}. Within a standard FE capacitor model,~\cite{Mehta} the depolarizing field in the ferroelectric film is proportional to $8\pi P\lambda_M/\epsilon d_{\rm FE}$, where $\lambda_M$ and $d_{FE}$ are the screening length of the metal electrodes and the thickness of the FE film. Within this model, the depolarizing field vanishes only if $\lambda_M \rightarrow 0$, in which case the surface polarization charges are fully compensated by the free charges in the metal. Fig.~\ref{pot_pto} shows a residual depolarizing field of about 0.06 V/\AA~  in the PTO layers that points in the direction opposite of polarization, indicating incomplete screening by the metallic electrodes. The residual field in the PTO superlattice is about 17\% of the depolarizing field calculated for an isolated PTO film. This ratio is comparable to $2\lambda_M / d_{PTO} \sim 13\%$ where the FE thickness $d_{\rm PTO}$ is 16 $\rm\AA$ and screening length of Fe $\lambda_M$ is about $1 \rm \AA$.~\cite{note_screening}  

\begin{figure}
\includegraphics*[width = 9cm]{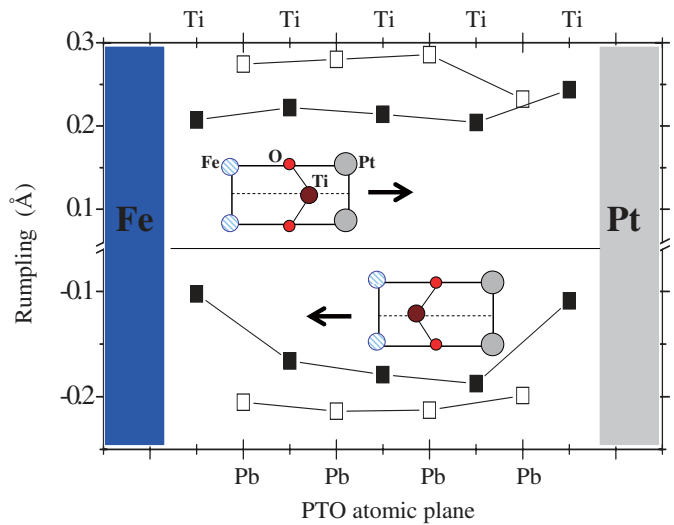}
\caption{ The rumpling parameter for each PbTiO$_3$ atomic plane in the (Fe)$_{11}$/(PbTiO$_3$)$_{9}$/(Pt)$_{8}$ superlattice. The insets illustrate the relative position of the interfacial TiO$_2$ plane with respect to the adjacent Fe and Pt atoms. The arrows mark the direction of the polarization in PbTiO$_3$. }
\label{pol}
\end{figure}

Using the Born effective charges~\cite{Zhong} and ionic displacements relative to the ideal positions, we estimate the polarization of the central unit cell to be about 75 and $70 \mu$C/cm$^2$ when the polarization points towards Pt and Fe, respectively. These values are slightly reduced compared to the calculated bulk value of $84\mu$C/cm$^2$. (For the BaTiO$_3$ based structure, we have found 20 and 18 $\mu$C/cm$^2$ for the respective polarization, whereas the bulk value is $25\mu$C/cm$^2$.)  Fig. \ref{pol} shows the rumpling parameter, defined as the relative displacement between the cations and oxygen ions for each atomic plane along the stacking direction of the PTO superlattice. The amplitudes of the interfacial rumpling are roughly 0.1~\AA~ higher when the polarization points towards Pt than when it points towards Fe. The higher amplitude in the former case is consistent with that observed in Pt/PbTiO$_3$/Pt~\cite{Sai}  and can be most likely attributed to the formation of metallic bonds between Pt and Ti at the Pt/PTO interface and the Fe-O bonding at the Fe/PTO interface. 

At a FE/metal interface, the ferroelectric polarization terminates and induces screening charges in the metal. The total screening induced charge density satisfies the Poisson equation
\begin{equation}
\dfrac{d^{2}V_{c}(x)}{d^{2}x}=-(e/\epsilon_{0})[\delta n^{\uparrow}(x)+\delta n^{\downarrow}(x)],
\label{Poisson}
\end{equation} 
where $V_c$ is the Coulomb screening potential in the metal and $\delta n^\sigma $ is the screening induced charge density of spin $\sigma = \uparrow, \downarrow$. If the metal is a FM, then the induced charges are spin dependent due to the exchange interactions and a local magnetization can be induced at the interface. The local induced magnetization and the Coulomb screening potential at a FM/dielectric~\cite{zhang} or FM/FE interface~\cite{cai} can be related through
\begin{equation}
\delta n^{\uparrow}(x)-\delta n^{\downarrow}(x)=-\frac{M_0}{1+JN_0} eV_{c}(x),
\label{screening}
\end{equation}
where $M_0$ and $N_0$ are the spontaneous magnetization and the total density of states, and  $J$ is the exchange splitting in the FM. Thus the induced magnetization switches sign when the orientation of the FE polarization switches. 

\begin{figure}
\includegraphics*[width=8.5cm]{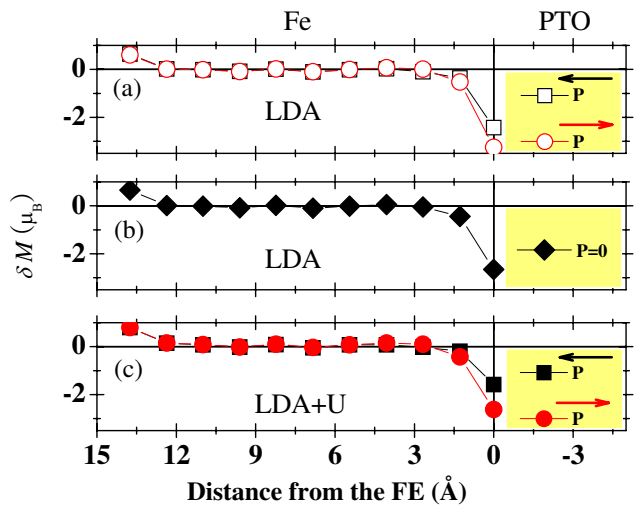}
\caption{ The layer by layer magnetic moment changes relative to the bulk of two Fe atoms per lateral unit cell of the (Fe)$_{11}$/(PbTiO$_3$)$_{9}$/(Pt)$_8$ superlattice. The PTO is in a (a) ferroelectric phase where the FE polarization directions are marked by the arrows and (b) paraelectric phase. (c)The magnetic moment changes calculated within LDA+U (where $U = 8$ eV).}   
\label{den_Fe}
\end{figure} 
To calculate the interfacially induced charge density of Fe due to the spin dependent screening, we project the density of states (DOS) of the Fe/PbTiO$_3$/Pt superlattice onto each atomic layer of Fe parallel to the interface. By integrating the DOS below the Fermi level separately for the spin-up and spin-down components, we obtain the layer by layer magnetization densities $n^\uparrow(x) - n^\downarrow(x)$ in the Fe unit cells.  In Fig.~\ref{den_Fe} (a), we plot the change of the layer by layer magnetization density with respect to the bulk value ($\sim4 \mu_B$/lateral unit cell counting two Fe atoms) for two opposite polarization directions in PTO. The total magnetization of Fe is significantly suppressed with respect to the bulk at the Fe/PTO interface and slightly enhanced at the Pt/Fe interface.  To understand this behavior, we plot $\delta M(x)$ of Fe when PTO is in a paraelectric (PE) phase,  i.e., $P=0$, in Fig.~\ref{den_Fe}(b) for comparison. It is clear that similar changes of the magnetic moment in Fe at both interfaces exist even in the absence of the FE polarization! We start by explaining this effect. 

\begin{figure}
\includegraphics*[width=8.5cm]{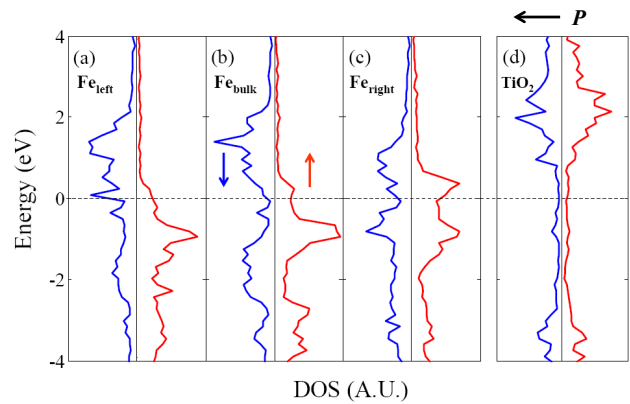}
\caption{The projected density of states (PDOS) of interfacial Fe atoms in the Pt/PTO/Fe superlattice at the (a) Pt/Fe interface,  (b) Fe bulk region,  and (c) Fe/PTO interface. Panel (d) shows the PDOS of the interfacial TiO$_2$ layer in PTO. The up and down arrows represent the spin majority and minority channels, respectively.}    
\label{dos}
\end{figure} 
The reduction of the iron's magnetic moment at the Fe/oxide interface is caused by charge transfer from Fe to the oxide caused by several mechanisms.  In Fig.~\ref{dos}, we show the DOS of Fe at both interfaces and compare them to that in the interior bulk region. At the Fe/PTO interface as shown in Fig.~\ref{dos}(c), the majority spin state at the Fermi level dominates, same as in bulk Fe. Thus losing charge from Fe to the oxide suppresses the majority spin more than the minority and leads to a reduction in the magnetic moment of Fe. The outflow of charge from Fe into the oxide is due to by two effects. First and the larger of the two is the formation of Fe-O bonds at the interface. Electronegative oxygen pulls the charge away from Fe. Second effect is filling of the so-called metal induced gap states (MIGS) that are clearly seen in Fig. \ref{dos}(d). This is universal for any metal/insulator interface when the metal Fermi level happens to be in the gap of the insulator. At the Fe/Pt interface, on the other hand, the density of states of Fe at the Fermi level is reversed with respect to the bulk as can be seen in Fig. ~\ref{dos} (a) and (b). Thus charge transfer from Fe into Pt due to the work function difference (the work function of Fe is 2 V less than Pt) results in losing more minority spins and therefore increases the magnetic moment in Fe. We find a similar effect  in our Pt/Fe bilayer calculation. The comparison with the $P=0$ case corroborates that the large moment change relative to the bulk is independent of polarization. In the following, we will remove this effect from the polarization dependent induced magnetization. 

Before proceeding, we make two relevant comments. The charge transfer from Fe to oxide is exacerbated by the reduction of the PTO band gap to about 2 eV within the LDA, compared to the experimental gap of 3.5 eV. By applying a Hubbard U correction to Ti $3d$ states, the band gap of the oxide is opened up\cite{lee08} and the overall magnetic moment reduction has been reduced as shown in Fig.~\ref{den_Fe}(c). However, the polarization dependence of the induced magnetization which is the main focus of the paper qualitatively stays the same.  Henceforth we only discuss the results from the LDA calculations.  We further note that a recent study~\cite{duan06} of a symmetric Fe/BaTiO$_3$/Fe junction reported  large induced magnetic moments on interfacial Ti atoms, even higher than the induced moments on the interfacial Fe atoms. The authors attributed the moment changes on Ti to a significant hybridization between the $3d$ orbitals of Ti and Fe at the interface. In contrast to the DOS in Ref.~\cite{duan06}, our interfacial Ti $d$ orbitals lie mostly above the Fermi level as shown in Fig. ~\ref{dos}(d). Therefore the moment changes on the interfacial Ti atoms are much smaller than those of Fe in our calculation and will be ignored in our study.

\begin{figure}
\includegraphics*[width=8.5cm]{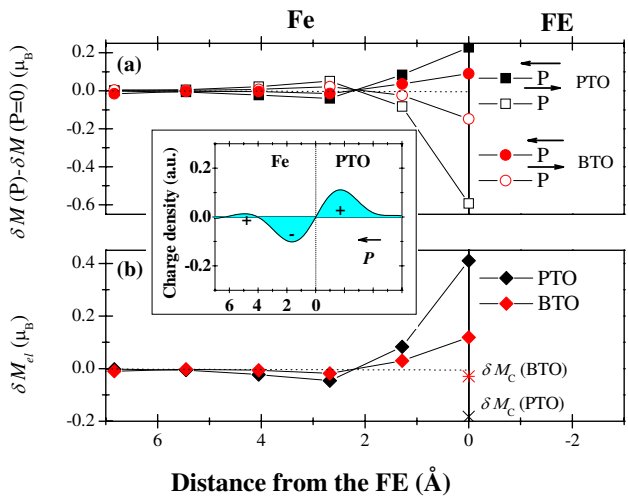}
\caption{(a) The magnitude of the polarization induced magnetization for two Fe atoms per lateral unit cell
vs. distance from the Fe/FE interface for the PTO and BTO based superlattice. Only half of the Fe cell close to the oxide is plotted.  (b) The electrostatic screening induced magnetization $\delta M_{el}$ (diamonds) and the magnetization induced by a chemical component related to polarization switching $\delta M_c$ (stars). The inset shows the induced charge density near the Fe/PTO interface.}   
\label{moment}
\end{figure}   
To focus on the polarization related changes of magnetization we subtract $\delta M(P=0)$ from $\delta M(\pm P)$ shown in Fig.~\ref{den_Fe}. Thus we obtain the polarization induced magnetization of Fe near the PTO interface for two opposite FE polarization directions as shown in Fig.~\ref{moment}(a).  We include the results from the BTO superlattice as well. In either case, the induced moments decay rapidly as function of distance away from the FM/FE interface, and change sign as the FE polarization direction switches. This suggests that if  bias is applied to switch the FE polarization, the magnetization becomes electrically controllable as consistent with our model calculation.~\cite{cai}  It is interesting to note that $\delta M$, however,  does not completely follow an exponential function as typically assumed in screening models. As the distance from the interface increases, the sign of $\delta M$ oscillates. We attribute this behavior to Fridel-like  oscillations in the metal.~\cite{Friedel} It is consistent with the induced charge density distribution (see the inset) calculated from the electrostatic potential near the interface, which also oscillates.   

Unlike the rest of the Fe layers,  the induced moment on the first interfacial layer of Fe is significantly asymmetric with respect to the direction of the polarization. To explain the asymmetry, we make a reasonable assumption that the total induced magnetization can be decomposed into two separate terms,  $\delta M_{el} $ and $\delta M_{c}$. The first term is related to electrostatic screening which changes sign as the polarization switches and the second term is an additional contribution which is related to the chemical bonding at the interface,
\begin{equation}
\begin{cases}
{\delta M(+P)}= {\delta M_{el}}+ {\delta M_{c1}} \\
{\delta M(-P)}= - {\delta M_{el}}+ {\delta M_{c2}}.
\end{cases}
\label{deltan}
\end{equation}
Since the opposite polarization directions result in different interfacial atomic configurations (Fe-O-Ti vs. Fe-Ti-O as illustrated in Fig.~\ref{pol} insets), $\delta M_c$ could be different in these equations. However, in an attempt to separate the effect of the electrostatic screening and chemical bonding, we assume the difference between $\delta M_{c1}$ and $\delta M_{c2}$  to be much less than the screening induced moment change and therefore $\delta M_{c1}\approx \delta M_{c2}$. We calculate $\delta M_P$ and $\delta M_c$ by subtracting and adding $\delta M(+P)$ and $\delta M(-P)$ and plot them in Fig.~\ref{moment}(b). Note that $\delta M_c$ exists only in the first interfacial layer. The magnitudes of both $\delta M_P$ and $\delta M_c$ are strongly dependent on the magnitude of the polarization in the FE oxide. 
In the PTO and BTO superlattice, the maximum magnitude of the screening induced moment change $\delta M_{el}(x)$ is about $0.4\mu_B$ and $ 0.13\mu_B$ per surface unit cell, corresponding to a magnetization density of $53$ and $14~\mu{\rm C/cm}^2$, respectively. The ratio between the two resembles closely the ratio between the calculated polarization of PTO and BTO. Thus the electrostatically induced magnetization is clearly linear in the FE polarization, in agreement with Eq. (1) and (2). This also suggests that our approximation for $\delta M_c$ is reasonable. On the other hand, the magnitudes of $\delta M_c$ are quite different, $0.19$ and $0.03$$\mu_B$, at the Fe/PTO and Fe/BTO interfaces, indicating a different interfacial bonding character between two different FEs with metals.~\cite{Stengel09}
 
To estimate the ME coupling, we use the dimensionless ratio between the screening induced magnetization density and bulk electric polarization of the FE,~\cite{rond} i.e., $\eta = \frac{\delta M_{el}}{P}$. The maximum values of $\eta$ are estimated to be 0.66 and 0.61 at the Fe/PTO  and Fe/BTO interface, respectively. For comparison, we also quantify the ME coupling using the definition that is strickly only suitable for bulk materials $\mu_0\delta M_{el} = \alpha E_c$, where $E_c$ is the coercive field of the FEs and $\mu_0$ is the vaccum permibility. Using a typical value of $E_c\sim100  {\rm kV/cm}$, we estimate the maximum interfacial magnetoelectric coefficient $\alpha$ to be $3 \times 10^{-10} {\rm G cm}^2 /V$ which is similar in order of magnitude as that reported by Duan {\it et al}.~\cite{duan06} However the coupling in our study arises solely from electrostatic screening, in contrast to the dominant contribution from orbital hybridization at the interface.~\cite{duan06} 

In summary, we have carried out a first principles  study of the interfacial magnetoelectric coupling in tri-component FM/FE/NM superlattices. Through a detailed analysis of the magnetic moment changes in Fe we have identified the major contributions to the interfacial magnetization. First of all, even in the absence of the FE polarization, the magnetic moment in Fe is significantly reduced by charge transfer from Fe to the oxide owing both to the Fe-O bond formation and metal induced gap states in the oxide. The latter effect is somewhat excerbated by the underestimated LDA band gap, that can be corrected at {\it e.g.}, the LDA+U level. Importantly, our LDA+U calculations show that once the FE polarization is turned on the difference between the magnetic moments for two opposite FE polarizations is not sensitive to the band gap. The polarization induced change of the magnetization at the interface has two separate contributions. We identify a "chemical" contribution that is asymmetric in FE polarization owing to the difference in the interfacial atomic geometry, and thus is interface specific, and the symmetric contribution arising from the spin dependent screening. The last effect is electrically controllable, linear in the FE polarization and leads to a substantial magnetoelectric coupling. 

This work is supported by the Office of Naval Research under grant N000 14-06-1-0362, NSF DMR-Career-0548182 grant. The authors acknowledge the Texas Advanced Computing Center (TACC) at the University of Texas at Austin for high perfermance computing resources. TC and QN were supported by DOE (DE-FG03-02ER45985), NSF(DMR0906025), Welch Foundation (F-1255), and NSFC (10740420252).


\begin{thebibliography}{10}
\bibitem{erenstein} W. Eerenstein, N. D. Mathur, and J. F. Scott, Nature (London) {\bf 442}, 759 (2006).
\bibitem{ramesh} R. Ramesh and N. A. Spaldin, Nature Materials {\bf 6}, 21(2007).
\bibitem{cheong} S.-W. Cheong and M. Mostovoy, Nature Materials {\bf6}, 13 (2007).
\bibitem{scott} J. F. Scott, Science {\bf 315}, 954 (2007).
\bibitem{zhang} S.F. Zhang, Phy. Rev. Lett. {\bf 83}, 640 (1999).
\bibitem{rond} J. M. Rondinelli, M. Stengel and N. A. Spaldin, Nature Nanotechnology {\bf3}, 46 (2007).
\bibitem{duan06} C. G. Duan, S. S. Jaswal, and E. Y. Tsymbal, Phys. Rev. Lett.  {\bf 97}, 047201 (2006)
\bibitem{cai} T.Y. Cai, S. Ju, J.K. Lee, N. Sai, Alexander A. Demkov, Q. Niu, Z.Y. Li, J.R. Shi, and E.G. Wang, Phy. Rev. B {\bf 80}, 140415 (R) (2009).
\bibitem{duan08} C.G. Duan, J.P. Velev, R.F. Saviririanov, Z.Q. Zhu, J.H. Chu, S.S. Jaswal, and E.Y. Tsymbal, Phys. Rev. Lett {\bf 101} 137201 (2008).
\bibitem{Fechner} M. Fechner, I. V. Maznichenko, S. Ostanin, A. Ernst,  J. Henk, P. Bruno,  I. Mertig, Phys. Rev. B {\bf 78}, 212406 (2008).
\bibitem{Niranjan} M. K. Niranjan, J. P. Velev, C. G. Duan, S. S. Jaswal, and E. Y. Tsymbal, Phys. Rev. B {\bf 78}, 104405 (2008).
\bibitem{Niranjan2}M. K. Niranjan, J. D. Burton, J. P. Velev, S. S. Jaswal, and E. Y. Tsymbal, Appl. Phys. Lett. {\bf 95} 052501 (2009).
\bibitem{zhe04} H. Zheng, J. Wang, S. E. Lofland, Z. Ma, L. Mohaddes-Ardabili, T. Zhao, L. Salamanca-Riba,
S. R. Shinde, S. B. Ogale, F. Bai, D. Viehland, Y. Jia, D. G. Schlom, M. Wuttig, A. Roytburd, R. Ramesh, Science {\bf303}, 661
(2004).
\bibitem{zav05} F. Zavaliche, H. Zheng, L. Mohaddes-Ardabili, S. Y. Yang, Q. Zhan, P. Shafer, E. Reilly, R. Chopdekar, Y. Jia, P. Wright, D. G. Schlom, Y. Suzuki, and R. Ramesh, Nano Letters {\bf5}, 1793 (2005).
\bibitem{VASP}  G. Kresse and J. Futhm{\"u}ller, Comput. Mater. Sci. 6, 15 (1996); Phys. Rev. B {\bf 54}, 11169 (1996).
\bibitem{exp} T. Mitsui, M. Adachi, J. Harada, T. Ikeda, S. Nomura, E. Sawgu- 
chi, and T. Yamada, in Landolt-B\"o rnstein Numerical Data and 
Functional Relationships in Science and Technology: Oxides, 
Landolt-B\"ornstein, New Series, Group III, Vol. 16, Ch. 1A (Springer-Verlag, Berlin, 1981). 
\bibitem{note_strain}The elastic strain exerted upon Fe due to the lattice mismatch with PTO yields only a 0.7$\%$  increase of Fe bulk magnetization.
\bibitem{Mehta} R. R. Mehta, B. D. Silverman, and J. T. Jacobs, J. Appl. Phys. {\bf 44}, 3379 (1973).
\bibitem{note_screening}The metal slabs on both sides are chosen much thicker than the screening length, as indicated by the zero electric field across the central region of the metal electrodes. 
\bibitem{Zhong} W. Zhong., R. D. King-Smith and D. Vanderbilt, Phys. Rev. Lett. 72 3618 (1994).
\bibitem{Sai} N. Sai, A.M. Kolpak and A.M. Rappe, Phys. Rev. B (R)  {\bf 72}, 020101 (2005).
\bibitem{Stengel09}  M. Stengel, D. Vanderbilt, and  N. Spaldin, Nature Materials {\bf 8} 392 (2009).
\bibitem{lee08} J.K. Lee and A.A. Demkov, Phys. Rev. B 78, 193104 (2008).
\bibitem{Friedel} J. Friedel, Nuovo Cimento (Suppl.) {\bf 7}, 287 (1958).

\end{thebibliography}
\end{document}